\documentclass[%
reprint,%
aps, prl, 
floatfix,
twocolumn,
superscriptaddress,
showpacs,
showkeys,
]{revtex4-1}

\usepackage{makeidx}
\usepackage{docs}%

\usepackage{ulem} 
\usepackage{color}

\usepackage{graphicx}
\usepackage{natbib}
\usepackage{bm}%
\usepackage[colorlinks=true,linkcolor=blue]{hyperref}%
\expandafter\ifx\csname package@font\endcsname\relax\else
 \expandafter\expandafter
 \expandafter\usepackage
 \expandafter\expandafter
 \expandafter{\csname package@font\endcsname}%
\fi
\makeindex

\begin{document}

\title{Melting artificial spin ice}%

\author{Vassilios Kapaklis} 
\affiliation{Department of Physics and Astronomy, Uppsala University, Box 516,
SE-75120, Uppsala, Sweden}%
\author{Unnar B. Arnalds} 
\affiliation{Department of Physics and Astronomy, Uppsala University, Box 516,
SE-75120, Uppsala, Sweden}%
\author{Adam Harman-Clarke}
\affiliation{Department of Physics and Astronomy, Uppsala University, Box 516,
SE-75120, Uppsala, Sweden}%
\affiliation{London Centre for Nanotechnology, University College London, 17-19 Gordon Street, London WC1H 0AH, UK}%
\affiliation{Universit\'{e} de Lyon, Laboratoire de Physique, \'{E}cole Normale Sup\'{e}rieure de Lyon, 46 All\'{e}e d'Italie, 69364 Lyon Cedex 07, France}%
\author{Evangelos Th. Papaioannou} 
\affiliation{Department of Physics and Astronomy, Uppsala University, Box 516,
SE-75120, Uppsala, Sweden}%
\author{Masoud Karimipour}
\affiliation{Department of Physics and Astronomy, Uppsala University, Box 516,
SE-75120, Uppsala, Sweden}%
\author{Panagiotis Korelis}
\affiliation{Department of Physics and Astronomy, Uppsala University, Box 516,
SE-75120, Uppsala, Sweden}%
\author{Andrea Taroni}
\affiliation{Department of Physics and Astronomy, Uppsala University, Box 516,
SE-75120, Uppsala, Sweden}%
\author{Peter C. W. Holdsworth}
\affiliation{Universit\'{e} de Lyon, Laboratoire de Physique, \'{E}cole Normale Sup\'{e}rieure de Lyon, 46 All\'{e}e d'Italie, 69364 Lyon Cedex 07, France}%
\author{Steven T. Bramwell}
\affiliation{London Centre for Nanotechnology, University College London, 17-19 Gordon Street, London WC1H 0AH, UK}%
\author{Bj\"{o}rgvin Hj\"{o}rvarsson}
\affiliation{Department of Physics and Astronomy, Uppsala University, Box 516,
SE-75120, Uppsala, Sweden}%


\begin{abstract}
Artificial spin ice arrays of micromagnetic islands are a means of engineering additional energy scales and frustration into magnetic materials. Here we demonstrate a magnetic phase transition in an artificial square spin ice and use the symmetry of the lattice to verify the presence of excitations far below the ordering temperature.  We do this by measuring the temperature dependent magnetization in different principal directions and comparing with simulations of idealized statistical mechanical models. Our results confirm a dynamical pre-melting of the artificial spin ice structure at a temperature well below the intrinsic ordering temperature of the island material. We thus create a spin ice array that has real thermal dynamics of the artificial spins over an extended temperature range.
\end{abstract}

\pacs{75.10.Hk, 75.78.-n}
\keywords{Artificial spin ice, Ising model, magnetization dynamics}

\maketitle

Geometric frustration is observed in many physical systems. The textbook example is the frustration of proton interactions in water ice, giving rise to proton disorder, as revealed in the pioneering experimental work of \citet{giauque} and the theoretical interpretation of  \citet{pauling}. Frustration in antiferromagnets analogous to the ice model was predicted theoretically by Anderson in 1956 \cite{Anderson}. Only much later was ice-type disorder observed in magnets, and most surprisingly, this was in ferromagnetic materials that were thus named `spin ice' \cite{Harris,Ramirez_nature,StevenTBramwell11162001}. The spin ice phenomenon relies on dipole-dipole interactions and the concept was later generalized to include arrays of magnetic islands with a spin ice type geometry, the artificial spin ices \cite{wang_nature, Lammert_NatPhys_2010, Ladak_monopoles, Mengotti:2010p941, cumings_nature, Schumann, Morgan:2010p980}. Beyond the ice-type systems, the property of frustration is responsible for the occurrence of thermodynamically metastable phases in a variety of systems, including structural (amorphous materials) \cite{Senkov20012183}, magnetic (spin glasses) \cite{PhysRevB.16.4630} and polymeric systems \cite{PhysRevLett.75.4638}. In general, the absence of unique ground states \cite{Lammert_NatPhys_2010} and the presence of kinetic constraints \cite{Glassy_dynamics} makes the study of excitations extremely important. These, in combination with the degree of frustration, are the driving forces for disorder in such systems \cite{fecht_nature}. An order-disorder transition has been proposed for colloid spin ice model systems with adjustable interactions \cite{PhysRevLett.97.228302} and dynamical studies have already proven the importance of distortions \cite{Colloid_Nature}. 

\begin{figure}
 \begin{center}
 \includegraphics[width=\columnwidth]{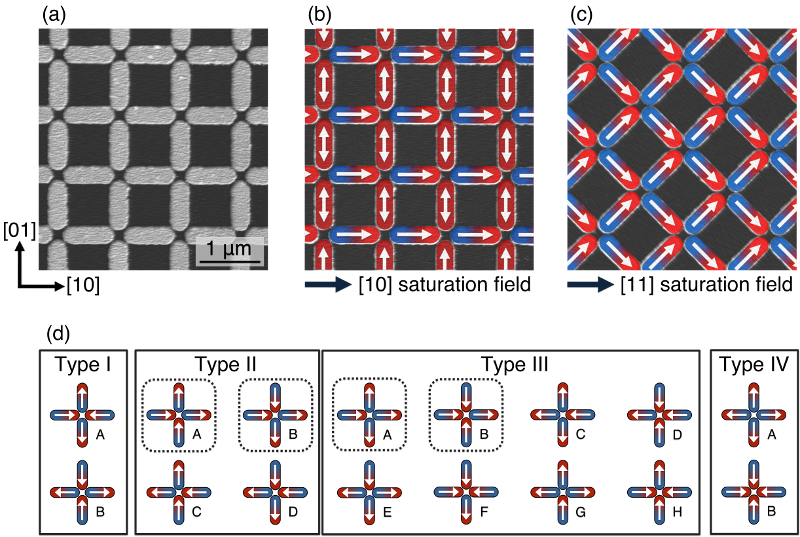}
  \end{center}
  \caption{(a) Atomic Force Microscopy (AFM) image of the artificial square spin ice array. The major symmetry axes are indicated in the lower left corner. (b) Remanent magnetic configuration of the spin ice array after applying a saturation field parallel to the [10] direction and (c) parallel to the [11] direction. (d) The 16 possible remanent magnetic configurations for the artificial square spin ice vertices. After applying a saturation field along the [10] direction there is a fourfold degeneracy of the possible remanent vertex configurations. In comparison, after removing the [11] field, there is only one possible remanent vertex configuration.}
\label{fig:AFM}
\end{figure}

The aim of the present work is to realize a thermally induced transition in an artificial spin ice array. In previous work on artificial magnetic spin ice, dynamics have been induced by changes in an applied magnetic field, rather than by thermal excitations \cite{Wang_Nisoli_JAP_2007, Nisoli_PRL_2007, PhysRevLett.101.037205, Nisoli_PRL_2010}. The use of field-cycling protocols has enabled the realization of quasi-degenerate spin states reminiscent of real spin ice \cite{wang_nature,cumings_nature}, as well as ordered states that are in static equilibrium \cite{Schumann}. All these results are obtained at temperatures far below the ordering temperature of the patterned structures. Recent work by Morgan \textit{et~al.} \cite{Morgan:2010p980} explored for the first time ground states in artificial spin ice, obtained during the growth of the pattern. 



Here we used $\delta$-doped Pd(Fe) to enable the exploration of the thermally driven order-disorder transition of the artificial spin ice structure. The Curie temperature of $\delta$-doped Pd(Fe) films can be adjusted by variation of the iron (Fe) layer thickness in the palladium (Pd) \cite{parnaste_jpcm,papaioannou_jpcm}. A 1.2 monolayer thick Fe layer was embedded between 10 monolayers of Pd, which resulted in a Curie temperature of $T_{\mathrm{C}}=230$ K. The film was patterned by an electron beam lithography and Ar ion milling into an artificial square spin ice array, of 750$\times$250 nm$^2$ islands with a period of 1000 nm, extending over a 1.5$\times$1.5 mm$^2$  area (Fig. \ref{fig:AFM}). The total magnetic moment for one island was determined to be $M_0 = 1.1\times10^{-16}$ Am$^{2}$ at 5 K. 


\begin{figure}
 \begin{center}
 \includegraphics[width=\columnwidth]{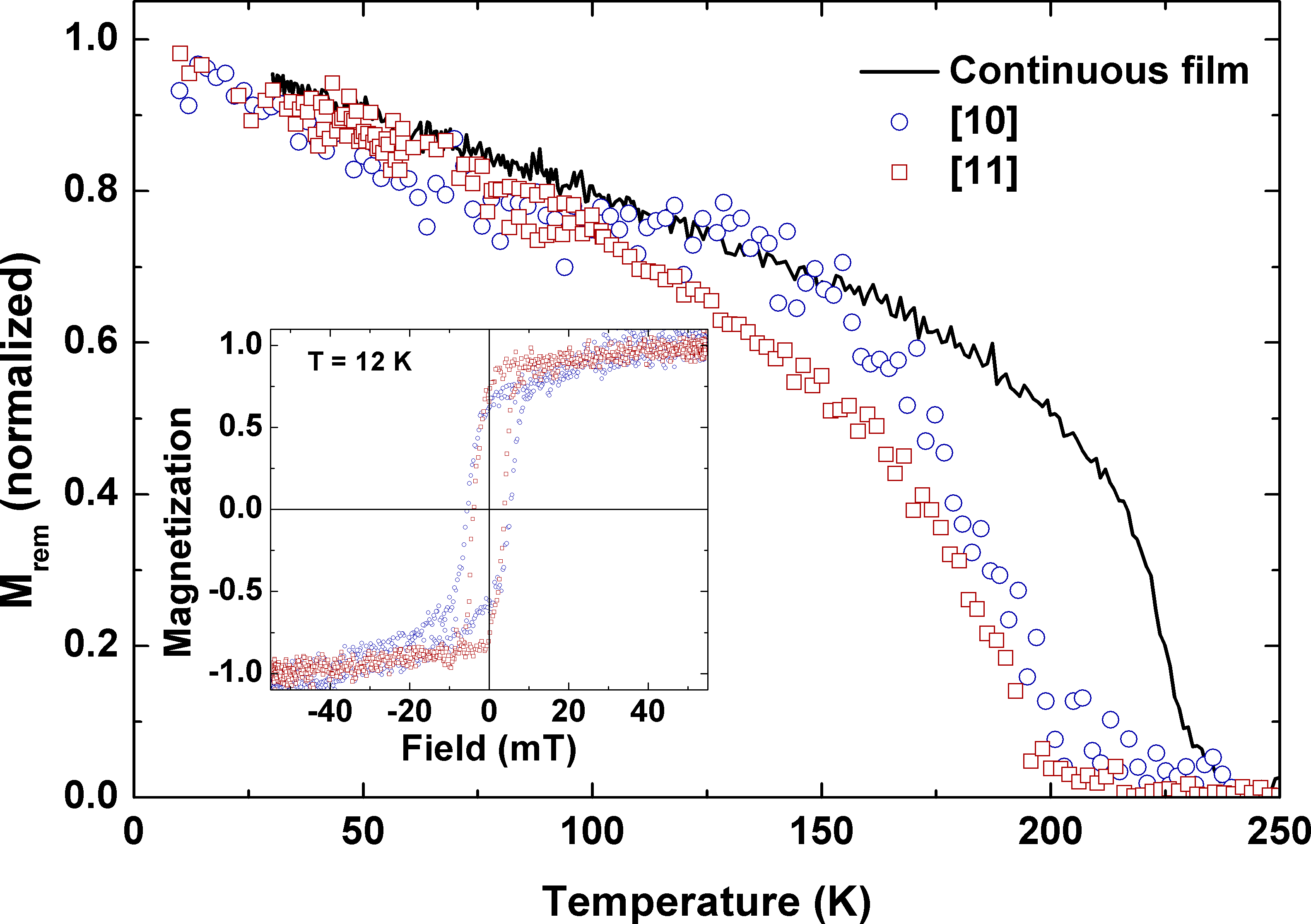}
  \end{center}
  \caption{Normalized remanent magnetization versus temperature curves after a magnetic field has been applied parallel to the [10] and [11] directions, compared with the remanent magnetization of the continuous film used for patterning the arrays. The collapse of the array magnetization at a temperature well below the Curie temperature of the material ($T_{\mathrm{C}}$ = 230 K) is consistent with the appearance of thermally induced dynamics of the macro-spins that comprise the array. The inset shows representative normalized magnetic hysteresis loops for a magnetic field applied parallel to the [10] and [11] directions, at a temperature of 12 K.}
\label{fig:mrem}
\end{figure}

Magnetic hysteresis loops  were recorded using the magneto-optical Kerr effect (MOKE) at temperatures from $5$ K to $300$ K \cite{papaioannou_jpcm} (see inset in Fig. \ref{fig:mrem}). The field was applied in two directions: (i) parallel to the [10] and (ii) parallel to the [11] directions of the patterned array, see Fig. \ref{fig:AFM}. When the field is removed, different populations of magnetic states for the different directions are obtained. For example, in the absence of thermal excitations, such a field cycling effectively removes vertices of type I, due to symmetry reasons. The normalized remanent magnetization, $M_{\mathrm{rem}}$, of the magnetic island array is compared to the continuous sheet material in Fig. \ref{fig:mrem}. The remanent magnetizations follow a similar trend up to $100$ K, but at higher temperatures the magnetic island array crosses over to a regime in which the magnetization is strongly reduced, approaching zero at around $200$ K, while that of the continuous film falls to zero at $T_{\mathrm{C}}= 230$~K.

In order to discuss the disordering of the magnetic island array, we distinguish between the microscopic magnetic moments within the islands (micro-spins, $\vec \sigma$) and the island macro-spins, composed of all $n$ moments within an island ($\vec{\cal{S}}={\displaystyle\sum\limits_{i=1}^n\vec \sigma_i / {|\displaystyle\sum\limits_{i=1}^n\vec \sigma_i|}}$). The coupling between the micro-spins is due to the exchange interaction, with a coupling constant $J$ related to the $T_{\mathrm{C}}$ of the magnetic material, while the coupling between the islands is magnetostatic. These interactions are long ranged but the characteristic energy scale can be defined through the interaction between macro-spins, $\vec{\cal{S}}$, on neighboring islands. The most relevant interaction is  between the nearest neighbours, coupling spins along rows or columns of the array. We define this coupling as $K= K_0 m^2$, where $K_0$ is of order $n^2$ and $m^2=\left<\left({1\over{n}}\displaystyle\sum\limits_{i=1}^n\vec \sigma_i\right)^2\right>$. 

 Introducing the thermally averaged quantity $m^2$ implies that thermal fluctuations of the macro and micro degrees of freedom are partially decoupled, allowing the two sets to order at different temperatures. This assumption could break down for smaller units, in which case the two ordering processes could be renormalized into a single transition, but it should be  valid for the islands studied here, consisting of $\sim 10^7$ micro-spins. Ordering of the macro-spins due to magnetostatic interactions should therefore occur at a temperature $T_{M} \leq T_{\mathrm{C}}$, as even in the limit $K_0 \gg J$, the term $m^2$ ensures that $K<J$ at $T_{\mathrm{C}}$.  Thus we expect disordering of the macro-spins at $T_{M}$ and a range of temperatures $T_{M} < T<T_{\mathrm{C}}$, in which the islands are decoupled.

The macro-spin $\vec{\cal{S}}$ is confined to lie along the long axis of the island through the magnetic shape anisotropy $K_{\mathrm{S}}$, which minimizes the collective dipolar energy of an island for the moment in one of the two configurations with Ising symmetry.  The energy barrier for reversing the direction of $\vec{\cal{S}}$ can be approximated by $E_r(T) =  K_{\mathrm{S}}V$, with $K_{\mathrm{S}}(T)=\mu_0 D (M_0 m)^2/2$, and is temperature dependent through $m^2$. Here $\mu_{0}$ is the permeability of free space, $D$ is the demagnetization factor of the island, and $V$ its volume. The active magnetic material is assumed to have a thickness of 1 nm \cite{PhysRev.67.351} when estimating $D$. The macro-spin reversal time scale can be assumed to follow a N\'eel-Arrhenius law $\tau =\tau_0 e^{E_r/(k_{\mathrm{B}}T)}$ where $k_{\mathrm{B}}$ is the Boltzmann constant and for inter-island magnetic fluctuations to occur, this barrier must be thermally accessible.

The temperature dependence of the effective moment of the islands is captured by using  $m^2 = (1-T/T_{\mathrm{C}})^{2\beta}$, with $T_{\mathrm{C}}= 230$ K, and $\beta=1/3$, which matches the temperature variation of the remanent magnetization of the continuous film shown in Fig. \ref{fig:mrem} to a good enough approximation for our purposes here. This leads to $E_r= E_0 m^2$, with $E_0/k_{\mathrm{B}} \approx 700$ K and $E_r/(k_{\mathrm{B}} T) =1$ at $195$ K, which ensures accessible thermal fluctuations for macro-spins below $T_{\mathrm{C}}$. 

The possible vertex configurations for artificial square spin ice are summarized in Fig. \ref{fig:AFM}(d). Two dimensional square ice does not have the same narrow band of low energy states as three dimensional spin ice, as the long range part of the dipolar interaction is less well screened than in the three dimensional counterpart \cite{PhysRevLett.95.217201}. As a result, the band of Pauling states cannot be considered quasi-degenerate at any level of approximation. The lowest energy configurations are made up of type I vertices, leading to an ordered ground state, with staggered magnetic moments. The type I vertices, which carry no moment, are effectively removed during field cycling and are therefore irrelevant in the absence of thermal excitations. The lowest energy states which both satisfy the ice rules and carry a magnetic moment are the subset of type II vertices. 
Hence the simplest model, that captures the essential physics observed in this multi-scale problem is a sixteen-vertex model in which the type I states are raised in energy, leaving a four-vertex (type II) manifold of ground states.

\begin{figure}
 \centering
 \includegraphics[width=\columnwidth]{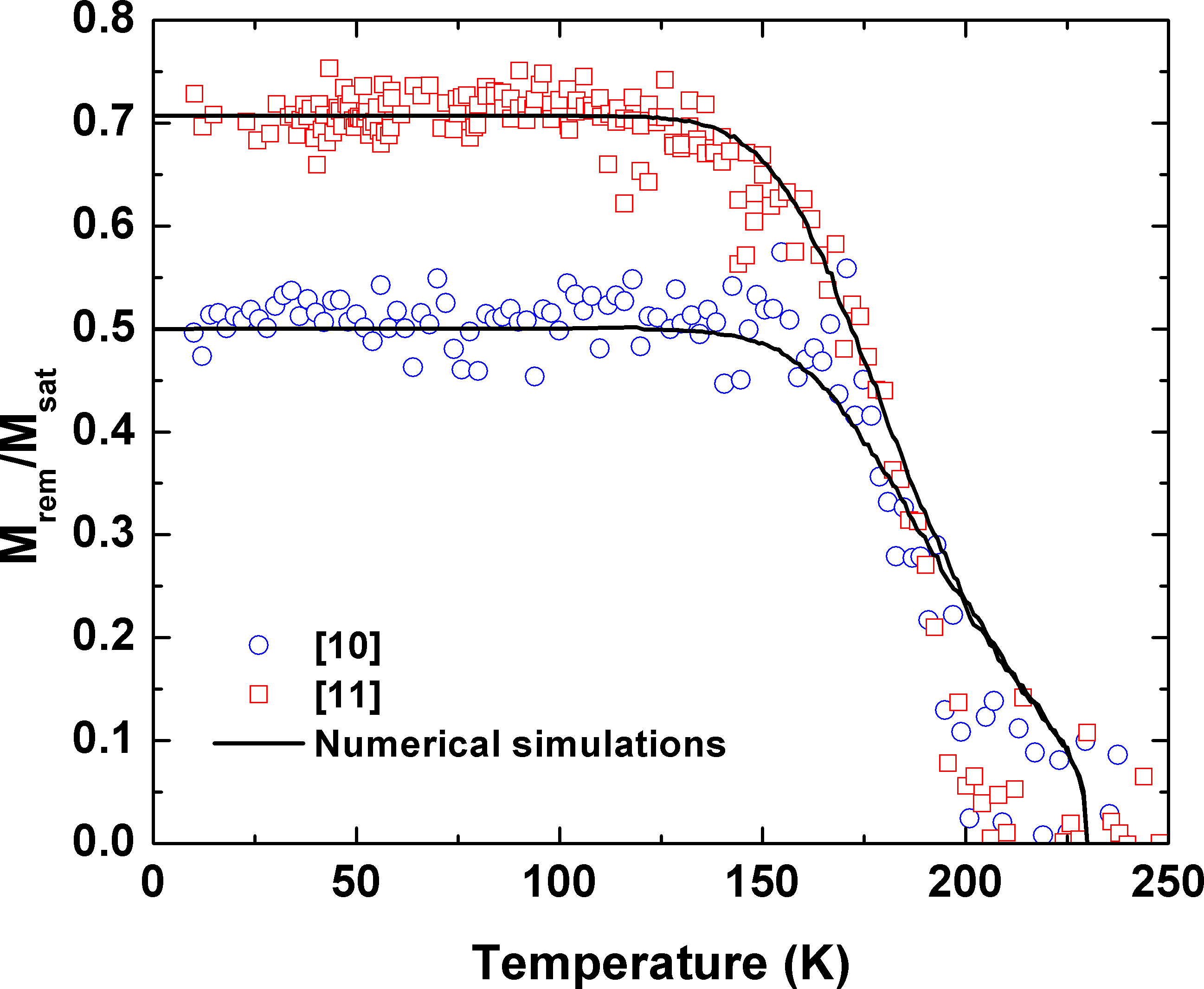} 
  \caption{The ratio of the remanent to saturation magnetization for the array.  At low temperatures the ratios are 0.5 and 0.707 for the [10] and [11] directions. This is consistent with an ordering of type II A + B  and type III A + B vertices as defined in Fig. \ref{fig:AFM}(d). Above 125 K, the macro-spin order diminishes (`pre-melts') which is well below the Curie temperature of the material ($T_{\mathrm{C}}$ = 230 K). The solid black lines are numerical simulations for the 16-vertex model, described in the text, that account for the thermally induced dynamics of the macro-spins.}
   \label{fig:ratio}
\end{figure}

Within this scenario the application of a saturation field along the [10] direction would result in a partially lifted degeneracy at low temperature, in favor of the ensemble of states shown in Fig. \ref{fig:AFM}(b). Applying the field along the [11] direction, followed by a return to remanence would lead to the selection of the fully ordered state illustrated in Fig. \ref{fig:AFM}(c). 
To capture the temperature dependence of the excitations we can therefore compare the ratio of the remanent and saturation magnetization, ${M_{\mathrm{rem}}}/{M_{\mathrm{sat}}}$, for these two directions.  The order parameter of the islands, $m^2$, cancels out, leaving the ordering behavior of the macro-spins $\vec{\cal{S}}$ only. The resulting ratio should therefore be $1/2$ for the [10] direction and $1/ \sqrt{2}$ for the [11] direction. As seen in Fig. \ref{fig:ratio}), the measured ratio falls on the prediction up to 125 K, illustrating collective behavior of the macro spins, with the single magnetic domain structure illustrated in Fig. \ref{fig:AFM}(b, c). Remarkably, the four vertex behavior, in which vertex defects can be neglected, remains up to temperatures where thermal fluctuations have reduced the squared order parameter of the islands, $m^2$, to less than one half of its zero temperature value. In the case of the [11] field, the data falls below the $1/ \sqrt{2}$ value at 125 K, while that for the [10] field stays on the line up to a slightly higher temperature. The magnetization, for the field applied in both directions, finally collapses as the ordering temperature of the material, $T_{\mathrm{C}}=230$ K, is approached.

\begin{figure}
   \centering
   \includegraphics[width=\columnwidth]{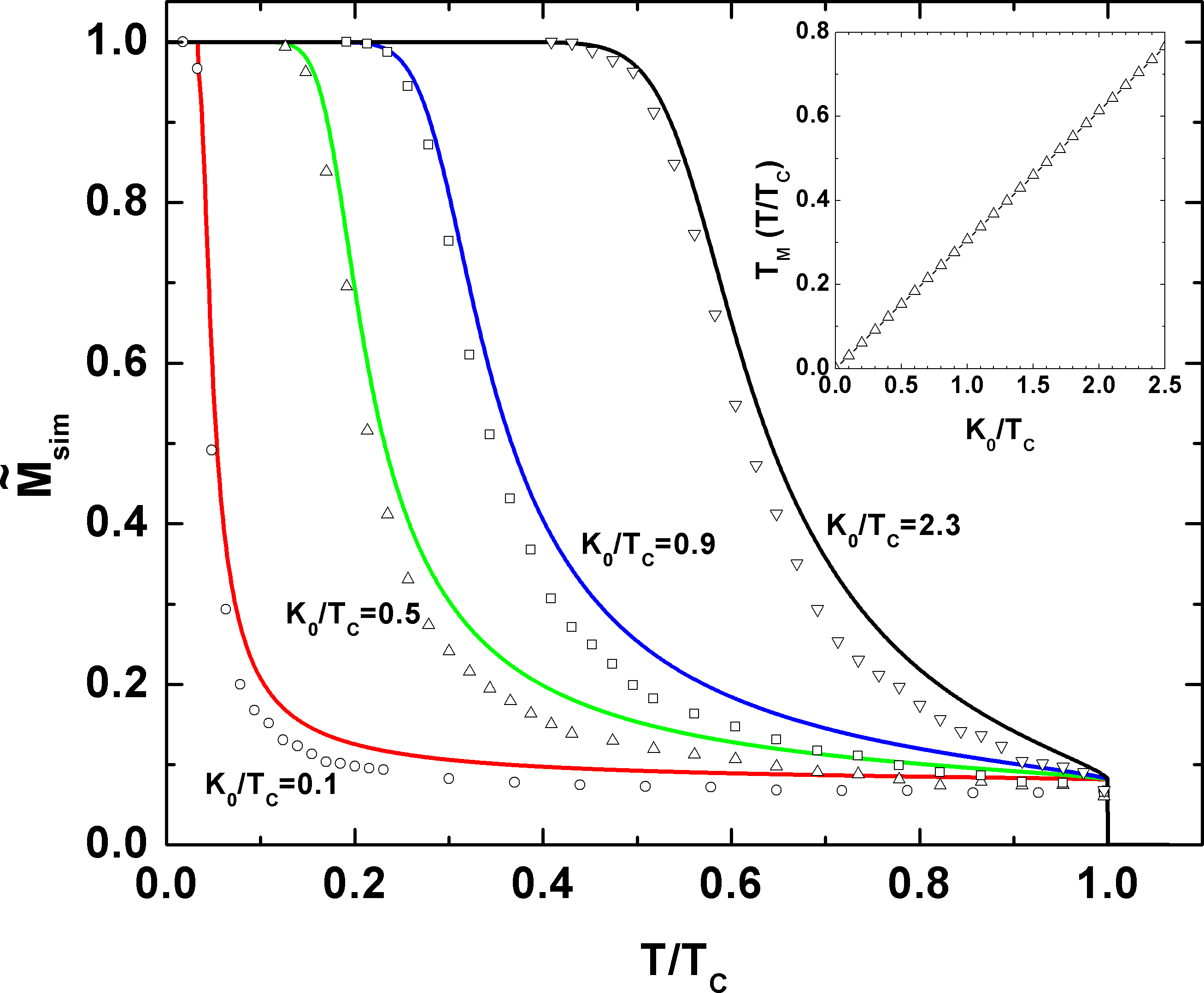} 
   \caption{Illustration of the thermal collapse of chain-like correlations between macro-spins. The array is approximated by an assembly of finite length Ising chains that order below a temperature, $T_{M}$. The points are numerical data from Monte Carlo simulations and the lines the analytical solution for the magnetization of an Ising spin chain of length $L=150$. The extent of this `finite size decoupling' regime can be varied by changing the coupling strength. Inset: The relation between the spin coupling constant and the pre-melting temperature $T_{\mathrm{M}}$, extracted from the Monte Carlo simulations. The finite size decoupling regime occurs for any value of the coupling $K_0$.}
   \label{fig:theory}
\end{figure}
We performed Monte Carlo simulations on a system with $N=45000$ macro spins, to model the experimental results. A sixteen-vertex ice model of $N$ Ising spins $\vec {\cal{S}}$ was used, arranged on a square lattice. To mimic the experimental conditions the energy of type I vertices was raised above those of type II and for simplicity we set  them  equal in energy to the type IV vertices, giving a single energy scale, $\epsilon_{III}-\epsilon_{II}=K$, $\epsilon_{IV}-\epsilon_{III}=K$. This model is equivalent to a grill of interpenetrating Ising spin chains coupled uniquely through the first neighbours along the chains [see Fig. \ref{fig:AFM}(d)]. The ground state is thus  an $\Omega=2^{2L}$ degenerate manifold of type II states with ferromagnetic ordering along chains (rows) of length $L=\sqrt{N/2}$ but with disorder among the chains. The parameter $K$ defines the effective coupling of macro-spins within a chain and its temperature dependence is obtained by scaling with $m^2$. 
As expected, there is no phase transition to an ordered state at finite temperature and in zero field the system remains disordered to $T=0$. However the model does have remarkable finite size effects related to the zero temperature critical point of the one dimensional Ising model \cite{ising,fsIsing}. This effect is illustrated in Fig. \ref{fig:theory}, which shows the temperature evolution of the magnetic moment for macro spins on a single  row (or column)  of spins, $\tilde{M}_{\mathrm{sim}} = |{1 \over L} \displaystyle\sum\limits^{L}_{i=1}\vec{\cal{S}}_{i}|$ for different values of $K_0/T_{\mathrm{C}}$. 
In each case the data are compared with $\tilde{M}_{\mathrm{ana}}$, calculated for a one dimensional Ising chain of $L$ macro-spins \cite{fsIsing}, with a coupling constant $K=K_0 m^2$, which is exact to leading order in $1/L$. The Monte Carlo results are excellently reproduced by the analytic expression for all values of $K_0/T_{\mathrm{C}}$, with small differences coming from corrections to scaling \cite{Hofgartner2011}. The chain moment approaches saturation below a crossover temperature, which we can interpret as $T_M$ when modeling the array. It scales logarithmically to zero with system size, thus for any reasonable system size, there is a finite region of temperatures for which each chain of the vertex model is ordered. 

This finite size chain ordering is at the origin of the observed pre-melting as it ensures that the field hysteresis protocol, leading to $M_{\mathrm{rem}}$, will order the chains giving a non-zero total moment for the vertex model below this crossover temperature scale. For a field along the [10] direction, the chains parallel to the field will be ordered but those perpendicular will be free, while for the field along [11] all chains should be ordered, as in the experiment. This field induced ordering is confirmed in Fig. \ref{fig:ratio}, where we compare calculated and experimental remanent to saturation magnetization ratio values. The numerical data was taken from simulations using two fitted parameters: $K_0 \approx 530$ K, corresponding to $K_0/T_{\mathrm{C}}=2.3$ (see Fig. \ref{fig:theory}), and a remanent applied field $B_0 = 1.8 \times 10^{-5}$ T. The simulation and experimental data are in good agreement up to around $200$ K, above which the experimental data falls more rapidly towards zero. This could arise due to the development of magnetic fluctuations perpendicular to the Ising spin axis or thermal macro-spin flips becoming more favorable. These are not taken into account numerically. 

The best fit value of $K_0$ compares favorably with calculated values: the direct interaction of the island magnetic poles  (treated as point charges) yields $K_{0} \approx 530$ K. In addition, this value of $K_0$ yields a crossover temperature, $T_M \sim 160K$ which is compatible with the experimental data.  The general agreement between experiment and simulation and the closeness of the fitted parameters to expected values therefore suggests that the sixteen vertex model proposed here, is indeed an accurate description of the experimental data. 

In both experiment and simulation, the crossover from the ordered spin ice regime to the pre-melting regime begins at a lower temperature for the [11] field direction than for the [10] direction. Pre-melting begins as vertex excitations disobeying the ice rules occur, breaking up the correlated chains of macro-spins. In such an excitation, a vertex of type II becomes one of type III through the flipping of the direction of a macro-spin. In the case of remanent magnetization along the [10] direction, there are vertex excitations that preserve the magnetic moment, for example type II $A+B$ $\rightarrow$ type III $A+B$, as illustrated in Fig. \ref{fig:AFM}(d). However, for $M_{\mathrm{rem}}$ along [11] all excitations are moment reducing. This suggests that the magnetization ratio will remain near the upper bound in the first stages of pre-melting for ordering along the [10] direction while the ratio will be reduced from this bound as soon as defects appear, for ordering along the [11] direction. These defects carry accumulations of effective magnetic charge and in the case of three dimensional spin ice, have been successfully described using the language of emergent magnetic monopoles~\cite{Castelnovo,Jaubert,Giblin}. In this case, the term `monopole' should be used with great caution, as long ranged interactions are less well screened for four-fold symmetry in a plane, than in spin ice. Here we have demonstrated the possibility to measure the order to disorder transition in artificial magnetic spin ice. The possibility of observing emergent monopoles is therefore conceivable, following the general approach which we describe in the design of spin ice arrays.

The authors acknowledge the support of the Swedish Research Council (VR), the Knut and Alice Wallenberg Foundation (KAW), the Swedish Foundation for International Cooperation in Research and Higher Education (STINT), the Icelandic Research Fund for Graduate Students and the Institut Universitaire de France (IUF). S.T.B. thanks the UK EPSRC for financial support. The patterning was performed at the Center for Functional Nanomaterials (CFN), Brookhaven National Laboratory, which is supported by the U.S. Department of Energy, Office of Basic Energy Sciences, under Contract No. DE-AC02-98CH10886. V.K., U.B.A. and B.H. thank Aaron Stein for support received during patterning at CFN. We thank Simon Banks {and Michel Gingras} for valuable discussions.


%

\end{document}